\documentclass[onecolumn, preprint, superscriptaddress]{revtex4-1}
\usepackage{hyperref}
\usepackage{graphicx}
\usepackage{amsmath}
\usepackage{wrapfig}
\usepackage{subcaption}
\usepackage{url}

\newcommand{\petsc}{\texttt{PETSc}}
\newcommand{\boutxx}{\texttt{BOUT++}}

\begin{document}
\title{Fluid simulations of plasma filaments in stellarator geometries with BSTING}

\author{B Shanahan}
\affiliation{Max-Planck Institut f\"ur Plasmaphysik, Teilinstitut Greifswald, Germany}
\author{B Dudson}
\author{P Hill}
\affiliation{York Plasma Institute, Department of Physics, University of York, Heslington, York YO10 5DD, UK}
\email{brendan.shanahan@ipp.mpg.de}

\begin{abstract}
 Here we present first results simulating plasma filaments in non-axisymmetric geometries, using a fluid turbulence extension of the \boutxx~framework. This is made possible by the implementation of the Flux Coordinate Independent scheme for parallel derivatives, an extension of the metric tensor components which allows them to vary in three dimensions, and development of grid generation.  Tests have been performed to confirm that the extension to three dimensional metric tensors does not compromise the accuracy and stability of the associated numerical operators. Recent changes to the FCI grid generator in \boutxx, including a curvilinear grid system which allows for potentially more efficient computation, are also presented.  Initial simulations of seeded plasma filaments in a non-axisymmetric geometry are reported.  We characterize filaments propagating in the closed-field-line region of a low-field-period, rotating ellipse equilibrium as inertially-limited by examining the velocity scaling and currents associated with the filament propagation.  Finally, it is shown that filaments in a non-axisymmetric rotating ellipse equilibrium propagate in a toroidally nonuniform fashion, and it is determined that the long connection lengths in the scrape-off-layer enable parallel gradients to establish, which has consequences for interpretation of experimental data.  

\end{abstract}
\maketitle

\section{Introduction}
Neoclassical transport is the dominant loss mechanism in sufficiently hot stellarator plasmas and can dominate in the plasma core~\cite{Ho1987}. In the outer, colder parts of the plasma, however, turbulence becomes more important and therefore dominates the plasma edge region~\cite{Helander2014}. Since the Wendelstein 7-X stellarator~\cite{Beidler1990} has been optimized to have low neoclassical transport, turbulent transport could become comparable to neoclassical losses even in the center of the plasma.  Wendelstein 7-X has already demonstrated novel edge physics; poloidally rotating filaments as measured by visible cameras~\cite{Kocsis2017}, and a high-frequency variation of limiter heat fluxes~\cite{Wurden2017} merit numerical investigation. Furthermore, the edge of Wendelstein 7-X in the island divertor configuration exhibits long connection lengths, such that cross field transport can become comparable to parallel transport.  Predicting this cross-field transport in high density, collisional, detached plasmas without an ad-hoc assumption for diffusion is a motivation of this work.  It is becoming increasingly important to simulate turbulence in non-axisymmetric configurations. 

In stellarator core plasmas, the most common method for simulating plasma turbulence is with gyrokinetic codes such as GENE~\cite{Gorler2011}, which is feasible due to the closed flux surfaces and the low collisionality.  However, the simulations are computationally expensive for long (on the order of confinement time) temporal and global spatial scales.  Additionally, GENE simulations are currently limited to flux-tube and flux-tube-ensemble geometries.

The high collisionality of tokamak and stellarator edge plasmas facilitates a fluid approach to turbulence simulations.  While there are several fluid turbulence simulation codes for tokamak geometries~\cite{Ricci2012, Tamain2014, Stegmeir2016}, previous attempts to develop such a simulation framework for stellarators have been unsuccessful.  

The recent implementation of the Flux Coordinate Independent (FCI)~\cite{Hariri2013} method for parallel derivatives in \boutxx~has allowed for simulations in non-axisymmetric geometries~\cite{ShanahanJPCS2016,Hill2017}. Instead of aligning the computational grid to magnetic field lines, the FCI method uses interpolation of field line mapping on poloidal (or, in the case of linear geometries, azimuthal) planes to obtain values for finite-difference differentiation parallel to the magnetic field.  In \boutxx, a cubic Hermite spline is utilized, although other methods have been implemented~\cite{Hill2017}.  The FCI method removes the inherent singularities in flux or field aligned coordinates around magnetic null points.  Additionally, since the computational grid is no longer aligned to the magnetic field, the simulation of complex geometries including X-points is possible.  For a more complete discussion of the FCI method, see References~\cite{Hariri2013,ShanahanJPCS2016,Hill2017}.

Here, we present the first results simulating plasma fluid turbulence in non-axisymmetric geometries, made possible by extensive modifications to the \boutxx~framework~\cite{Dudson2009, Dudson2015}. Section~\ref{sec:changes} describes the recent modifications to the \boutxx~framework which are relevant for this work.  Initial testing of the modified framework is described in Section~\ref{sec:testing}, where Sections~\ref{sec:diffusion} and~\ref{sec:mms} test the accuracy FCI parallel gradient operators and their associated boundary conditions, and Section~\ref{sec:laplacianinversion} reports the modifications to the Laplacian inversion algorithms. Section~\ref{sec:poloidalgrids} introduces a new curvilinear coordinate system for FCI simulations in~\boutxx~which is used in Section~\ref{sec:filaments} to simulate plasma filaments in non-axisymmetric geometries; filaments in the closed-field-line region of a rotating ellipse geometry are determined to be inertially-limited and exhibit a toroidally non-uniform propagation, a result which has implications for interpretation of experimental data.  Finally, Section~\ref{sec:w7xgrids} describes how the curvilinear FCI grids can be used for simulation of realistic geometries, namely Wendelstein 7-X.     

\subsection{Modifications to the \boutxx~framework}
\label{sec:changes}
The \boutxx~framework is a modular, object oriented and open source framework for fluid simulations with an international team of developers~\cite{Dudson2009}.  This paper presents recent progress in modifying \boutxx~to Simulate Turbulence In Non-axisymmetric Geometries under the ``BSTING'' project.

Previous work in simulating non-axisymmetric geometries has focused on the conventional \boutxx~framework, which is a 3D code but was written with metric tensor components which vary in two dimensions due to an assumption of toroidal symmetry.  For an accurate simulation of plasma dynamics in stellarators, BSTING must include metric components which are fully three dimensional.  This extension to three dimensions is simple in principle (and was in fact mentioned in the introduction of the original \boutxx~paper~\cite{Dudson2009}), but unfortunately the geometrical components are integral to many different parts of the code, and the work presented here has required extensive modifications to the framework.

The majority of modifications are primarily focused on the numerical methods of spatial operators and do not affect file handling, parallelization, post processing, and many other functions in \boutxx.  Development has focused on implementing operators relevant to edge transport and turbulence simulations: spatial derivatives of scalar fields which vary in three dimensions, and Laplacian inversion.  Here we address the most relevant issues: the accuracy of spatial gradient operators, boundary condition implementation, and Laplacian inversion which allows plasma potential to be calculated from vorticity.  The following section provides initial tests for the implementation of these methods.

\section{Testing}
\label{sec:testing}
The development of BSTING is an extensive modification to the \boutxx~framework, and therefore careful testing of numerical accuracy is required.  In this section, we concentrate on ensuring the accuracy of spatial derivatives, boundary conditions, and Laplacian inversion.  All tests in this section use a geometry where the poloidal planes are described by the radial x-coordinate and vertical z-coordinate while the y-coordinate describes the toroidal (or longitudinal in linear geometries) direction.  The FCI operators therefore interpolate the relevant values based on field line mapping in the x-z planes.   In Section~\ref{sec:poloidalgrids} we will discuss an alternative coordinate system for complex geometries.  

\subsection{Flux surface mapping using heat diffusion}
\label{sec:diffusion}
\begin{figure}[h!]
\centering
\begin{minipage}{16pc}
\includegraphics[width=6.0cm]{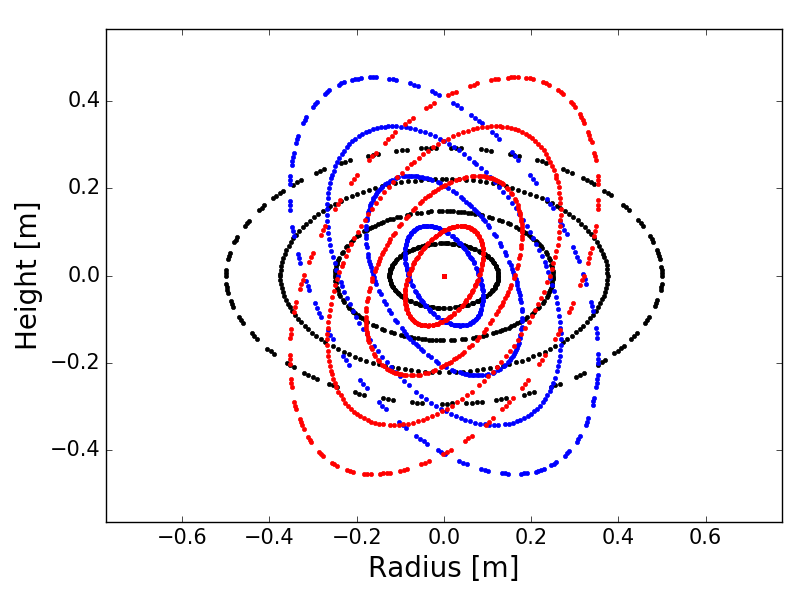}
\caption[]{Poincare plot indicating the flux surfaces in the analytic straight rotating ellipse equilibrium as calculated by the Zoidberg grid generator.}
\label{fig:stellsurfaces}
\end{minipage}\hspace{2pc}%
\begin{minipage}{16pc}
  \includegraphics[scale=0.08]{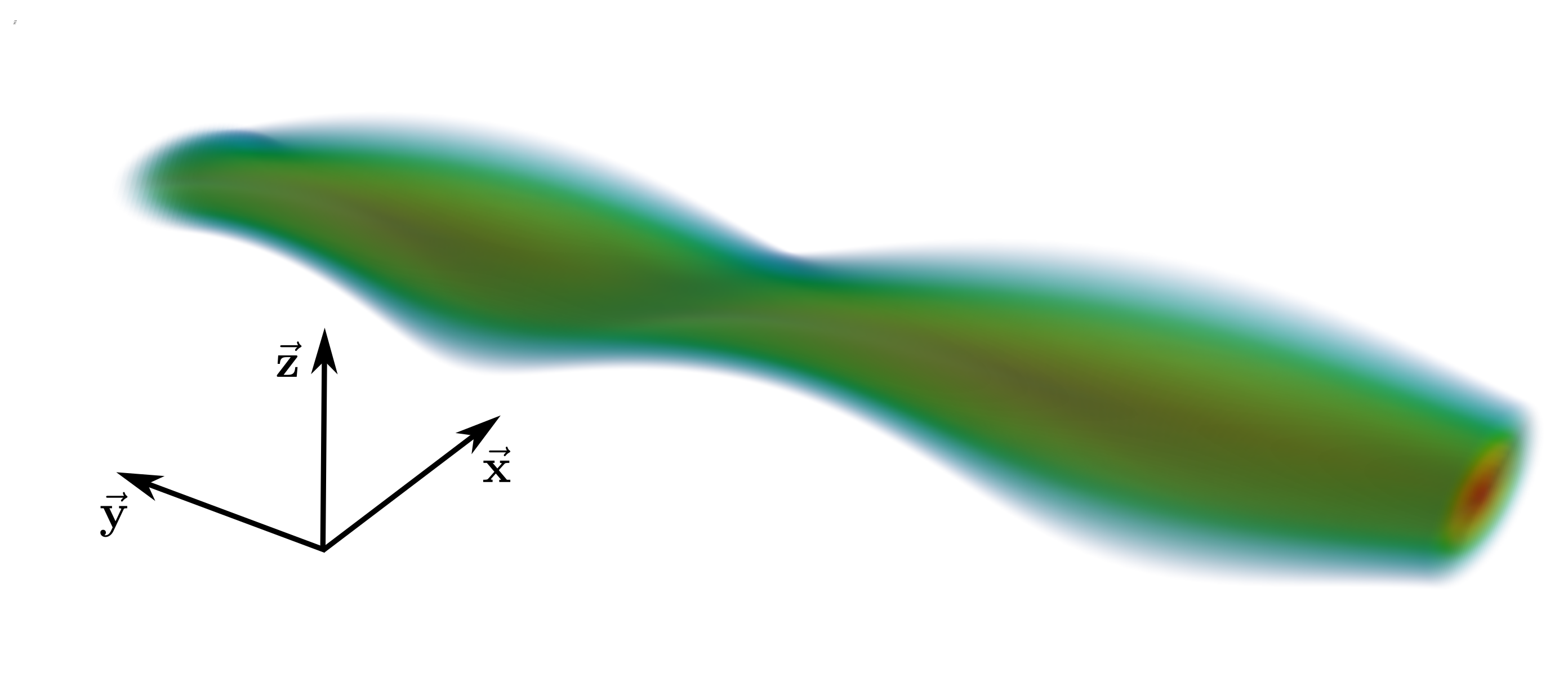}
\caption[]{Flux surfaces for a straight rotating ellipse equilibrium as calculated using the Flux Coordinate Independent operators in BSTING, reproducing to the test shown in Figure 4 from Reference~\cite{Shanahan2016}} \label{fig:stellarator_flux}
\end{minipage} 
\end{figure}

A potential issue with the implementation of the FCI scheme as discussed in section~\ref{sec:changes} is that since the poloidal planes are not orthogonal to the magnetic field lines, there could be a considerable pollution of perpendicular dynamics due to the projection of parallel effects~\cite{Gunter2005}.  A simple and common test to ensure the proper calculation of parallel dynamics using the FCI method in complex geometries is to implement a parallel diffusion model such as that shown in Equation~\ref{eq:diffusion}.  

\begin{equation}
\label{eq:diffusion}
\frac{\partial f}{\partial t} = \nabla \cdot \left({\bf{b}}{\bf{b}} \cdot \nabla f \right) \equiv \nabla^2_\parallel f
\end{equation}
where $\textbf{b}$ is the magnetic field vector.
Here the diffusion model in Equation~\ref{eq:diffusion} is used to test the numerical diffusion in a rotating ellipse equilibrium as done in References~\cite{Hill2017,ShanahanJPCS2016}.  Specifically, we will simulate this model on a rotating ellipse geometry, the flux surfaces of which are shown in Figure~\ref{fig:stellsurfaces}.  

\begin{figure}[htbp!]
\centering
\includegraphics[scale=.5]{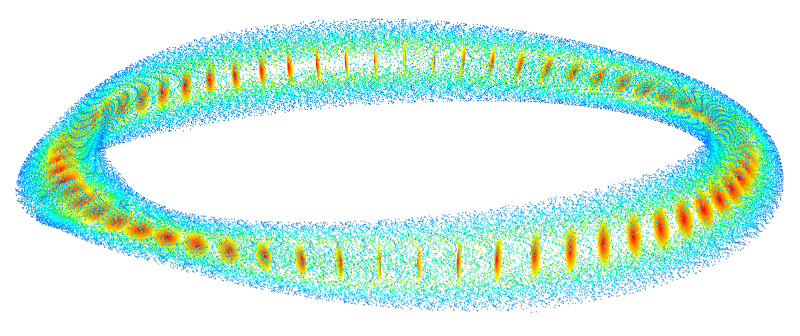}
\caption[]{Non-axisymmetric flux surfaces for a toroidal rotating ellipse equilibrium as calculated in BSTING.}
\label{fig:re_flux}
\end{figure}

Figure~\ref{fig:stellarator_flux} illustrates that simulating a parallel diffusion model qualitatively reveals the flux surfaces for a rotating ellipse equilibrium, recovering the results from~\cite{ShanahanJPCS2016,Hill2017} -- however this result differs in that it uses fully three dimensional metric tensor components, whereas the previous results utilized a metric tensor that varied in only two dimensions.  This added flexibility also allows for non-axisymmetric toroidal geometries.  Figure~\ref{fig:re_flux} indicates the flux surfaces as calculated by BSTING in a toroidal rotating ellipse geometry.  The red surfaces indicate the 2D projection on each poloidal plane, and the blue/green cloud is the interpolated function between the poloidal planes.  

This heat flux mapping indicates that the FCI operators are capable of simulating non-axisymmetric geometries after the transition to three dimensional metric tensors in BSTING.  The following section will use a more quantitative method to ensure the numerical operators and implementation of boundary conditions with three dimensional metric tensors have sufficiently small numerical error.  

\subsection{Method of manufacturing solutions for parallel derivatives}
\label{sec:mms}
\begin{figure}[htbp!]
\centering
\includegraphics[scale=.5]{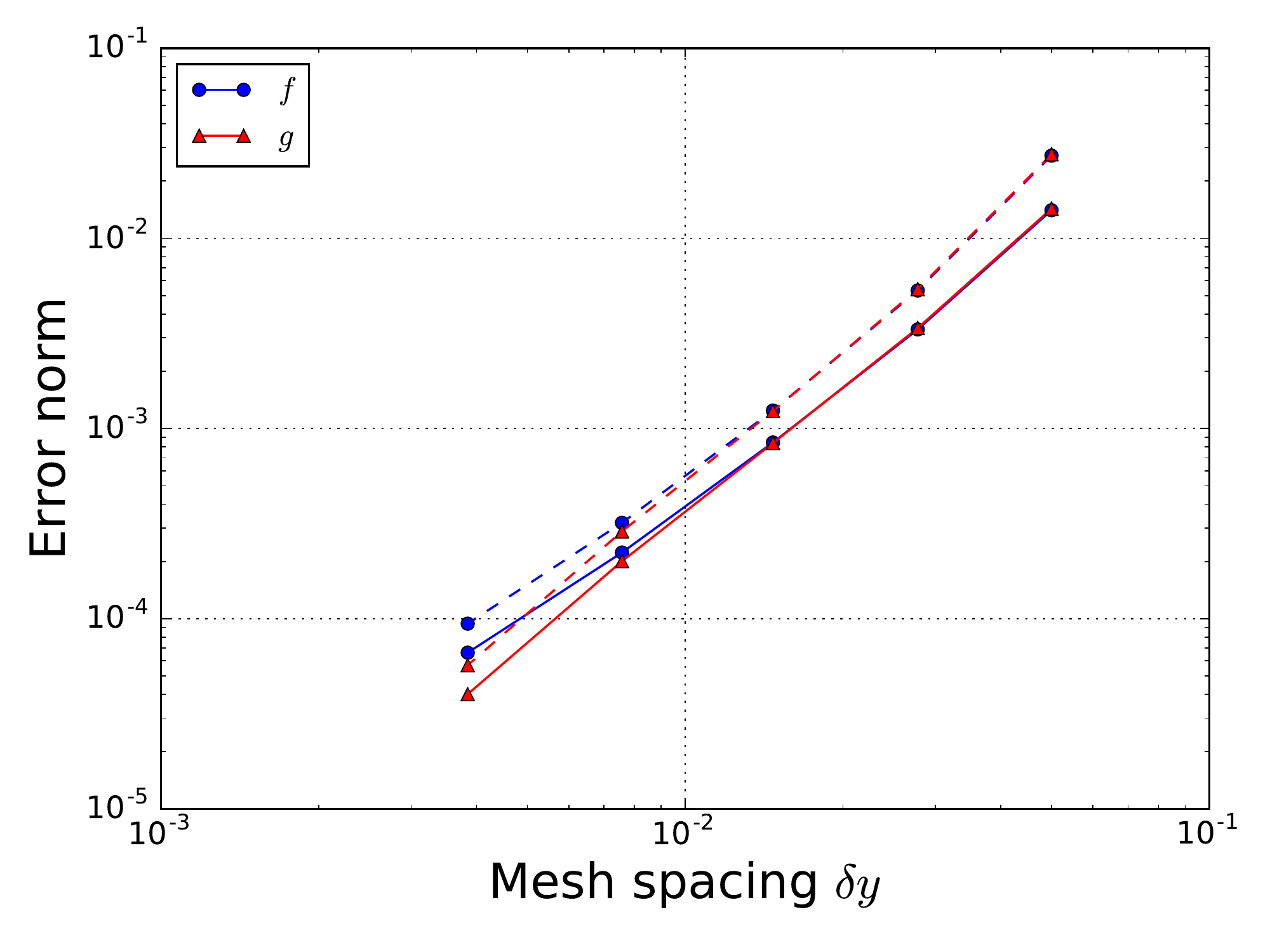}
\caption[]{Second order convergence for FCI operators in BSTING: the slope of the fits are 2.06 and 2.26 for $f$ and $g$, respectively.  The dashed lines indicate the maximum error (described as $l_\infty$ in Reference~\cite{Hill2017}).}
\label{fig:MMStesting}
\end{figure}

Imposing correct boundary conditions on plasma fluid turbulence simulations is complicated~\cite{Loizu2012} -- but the FCI method has particular issues at the boundaries, since the field lines can leave the domain before reaching the next toroidal plane, therefore leading to non-uniform grid point spacing for interpolation and complicating the correct calculation of derivatives.  There have been a few recent advances in boundary condition calculation for FCI operators; \boutxx~utilizes the Leg-Value-Fill (LVF) method detailed in Reference~\cite{Hill2017}.
In this section we extend previous testing~\cite{Hill2017} using the Method of Manufactured Solutions~\cite{Salari2000,Roache2002} to ensure that the extension to three dimensional metric tensors has not diminished the accuracy and stability of the framework.  Two coupled differential equations were therefore simulated for a single time step:

\begin{equation}
\label{eq:MMS1}
\frac{\partial f}{\partial t} = \nabla_\parallel g + D \nabla^2_\parallel f
\end{equation}
\begin{equation}
  \label{eq:MMS2}
\frac{\partial g}{\partial t} = \nabla_\parallel f + D \nabla^2_\parallel g
\end{equation}
where parameters are identical to those in Reference~\cite{Hill2017}; namely, $D$=10, and the domain measures 0.1 x 10 x 1 (x,y,z) meters.  The magnetic geometry is a sheared slab, such that ($B_x$,$B_y$,$B_z$) = (0,1,0.05 + (x-0.05)/10). The manufactured solutions are also those from Reference~\cite{Hill2017}:

\begin{equation}
  \label{eq:MMS3}
  f = \sin \left(\bar{y} - \bar{z}\right) + \cos(t)\sin \left(\bar{y} - 2\bar{z}\right) 
\end{equation}
\begin{equation}
  \label{eq:MMS4}
  g = \cos \left(\bar{y} - \bar{z}\right) - \cos(t)\sin \left(\bar{y} - 2\bar{z}\right) 
\end{equation}
where $\bar{y}$ and $\bar{z}$ are normalized between 0 and $2\pi$.  The diffusion terms in Equations~\ref{eq:MMS1} and~\ref{eq:MMS2} scale with y-spacing, and do not affect the convergence of $\nabla_\parallel$. Therefore the grid is scaled in y and z simultaneously.  Figure~\ref{fig:MMStesting} indicates the convergence of FCI operators in BSTING, including LVF boundary conditions.

Figure~\ref{fig:MMStesting} indicates a second order convergence of our operators.  Explicitly, the convergence order is 2.08 for $f$, and 2.26 for $g$.  A second order convergence is expected, as the FCI operators are second-order-central-differencing operators.

Having established the accuracy and stability of the FCI operators and the associated LVF boundary conditions in BSTING, the following section describes the implementation of Laplacian inversion routines which allow for the calculation of plasma potential from vorticity.  

\subsection{Laplacian inversion with complete poloidal metrics}
\label{sec:laplacianinversion}
One of the advantages of \boutxx~is its modular nature; numerical methods can be modified without compromising the stability or accuracy of the rest of the framework.  For this reason, several different methods for Laplacian inversion have been implemented in \boutxx.  Unfortunately for BSTING, many of these routines assume a periodicity in one direction (the z coordinate, usually the toroidal angle in tokamak simulations), since \boutxx~was originally designed to simulate turbulence in tokamak scrape-off-layers.  Recent work on implementing the Hermes model~\cite{Dudson2016} in \boutxx~has included several new numerical methods.  One of these is the implementation of a Laplacian inversion routine in three dimensions, which inverts an inhomogeneous Helmholtz equation in the conservative form:

\begin{equation}
  \label{eq:Laplacian}
  \nabla \cdot \left(A \nabla_\perp f \right) + Bf = b
\end{equation}
where $A$ and $B$ are coefficients set based on the equation to be solved, $b$ is most often vorticity and $f$ is the unknown quantity for which one solves (usually plasma potential). In most cases for fluid turbulence simulations, $B=0$ so that this equation becomes a Laplacian equation.  Here, the Laplacian is solved at each poloidal or azimuthal slice separately.  The discretization of Equation~\ref{eq:Laplacian} is then described in terms of fluxes through cell faces in the poloidal plane:

\begin{align}
  \label{eq:lapinv}
  \begin{split}
  \frac{1}{J}\frac{\partial}{\partial x}\left(JAg^{xx}\frac{\partial f}{\partial x}\right) &+ \frac{1}{J}\frac{\partial}{\partial z}\left( JAg^{zz}\frac{\partial f}{\partial z}\right) \\ &+ \frac{1}{J}\frac{\partial}{\partial x}\left( JAg^{xz}\frac{\partial f}{\partial z}\right)  + \frac{1}{J}\frac{\partial}{\partial z}\left( JAg^{xz}\frac{\partial f}{\partial x}\right) + B f = b
  \end{split}
\end{align}
where $J$ is the Jacobian, $g^{ij}$ are the metric tensor components, and $A$, $B$ and $b$ are variables which are specific to each situation -- for instance $b$ is often vorticity in plasma turbulence simulations. The current implementation of this solver utilizes the \petsc~suite of data routines~\cite{Balay1997}, which is available with several features including preconditioners for efficient computation. This implementation differs from conventional \boutxx~since it includes the off-diagonal metric terms ($g^{xz}$).  By setting the metric tensor components, $g^{ij}$, to non-zero values and comparing the implemented inversion routine using \petsc~to explicit calculation of Equation~\ref{eq:lapinv} indicated a difference of less than $10^{-15}$.  Testing with zero-value diagonal metric tensor components indicated similar errors relative to the implementation without off-diagonal metrics in \boutxx, suggesting proper convergence of the inversion routines.  

Having implemented the FCI operators and Laplacian inversion with Cartesian poloidal grids, the BSTING project is now capable of simulating turbulence in non-axisymmetric geometries.  A significant challenge for this method, however, is to handle the entire plasma cross section in a Cartesian poloidal grid while neglecting the plasma core and far edge.  One solution to this issue is to use a penalization function to mask the areas where the variables should not be evolved.  This method has been used previously in \boutxx~\cite{Shanahan2016} to remove solid-density magnetic coils in the simulation domain and is currently used in with FCI operators in GRILLIX~\cite{Stegmeir2016} to mask the plasma core and far scrape-off-layer.  The disadvantage of this method is that it requires a large poloidal grid for a relatively small computational area.  In the following section we present a new method for generating FCI grids in \boutxx~and BSTING which does not use a grid over the entire plasma cross section, potentially providing faster computation.  

\section{Elliptic FCI Grid generation}
\label{sec:poloidalgrids}
\subsection{Implementation of Elliptic Grids}
\begin{figure}[htbp!]
\centering
\includegraphics[scale=.45]{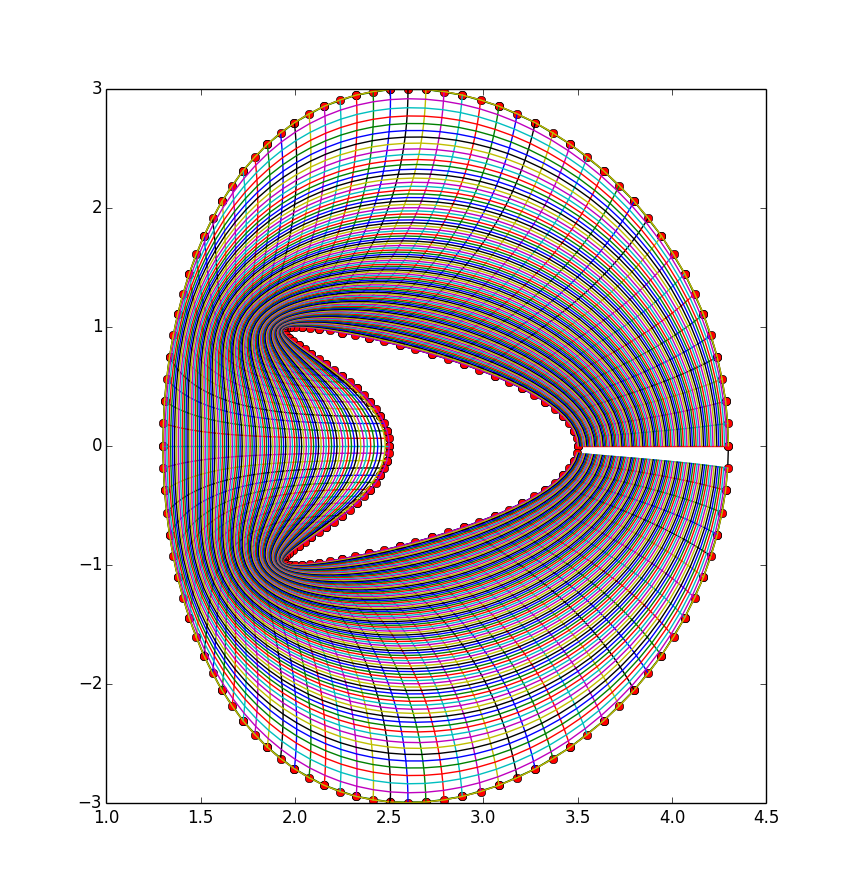}
\caption[]{An example of a curvilinear grid generated by the Zoidberg grid generator, which can be found in the \boutxx~manual~\cite{BOUTmanual}.}
\label{fig:pol_grid_example}
\end{figure}

While all previous simulations using the FCI method have used poloidal planes with Cartesian coordinates~\cite{ShanahanJPCS2016,Hill2017,Stegmeir2016,Hariri2013,Hariri2014}, this is not required.  The method is independent of the poloidal grid system as long as interpolation in these planes is correctly calculated and communicated. Here we present recent results using structured, non-Cartesian poloidal grids which are still logically rectangular~\cite{Thompson1982,Thompson1985}.  As an illustration of this method, Figure~\ref{fig:pol_grid_example} illustrates a sample grid with independent inner and outer surfaces.

These new grids have been added to the \boutxx~FCI grid generator, Zoidberg, and are included in a recent release of \boutxx~(version 4.1).  These grids are particularly advantageous as they include a periodic direction which could potentially increase computational efficiency. A grid is generated by prescribing an inner and outer surface, and then inverting an elliptic equation to connect the inner and outer points.  Both the inner and outer surface shapes are independently prescribed, and can be described using various methods: Zoidberg includes an flux surface shape generator, which will describe a shape based on elongation, triangularity and indentation.  Alternatively, one can use the Zoidberg field line tracer to construct flux surfaces from a given magnetic field (i.e. from VMEC, a vacuum field solver, or an analytic magnetic field description), and generate a shape based on this flux surface mapping.  

These grids provide an additional degree of flexibility and avoid some potential problems -- primarily how to mask the core/outer edges:  perpendicular (poloidal) boundaries are logically perpendicular to the grid cells, simplifying the imposition of boundary conditions -- although parallel boundaries must still utilize a method such as the Leg-Value-Fill method~\cite{Hill2017} discussed earlier. Some minor modifications to numerical operators are required for this poloidally-curvilinear coordinate system, which are discussed in the Appendix of this work.  

\begin{figure}[htbp]
\centering
\includegraphics[scale=.6]{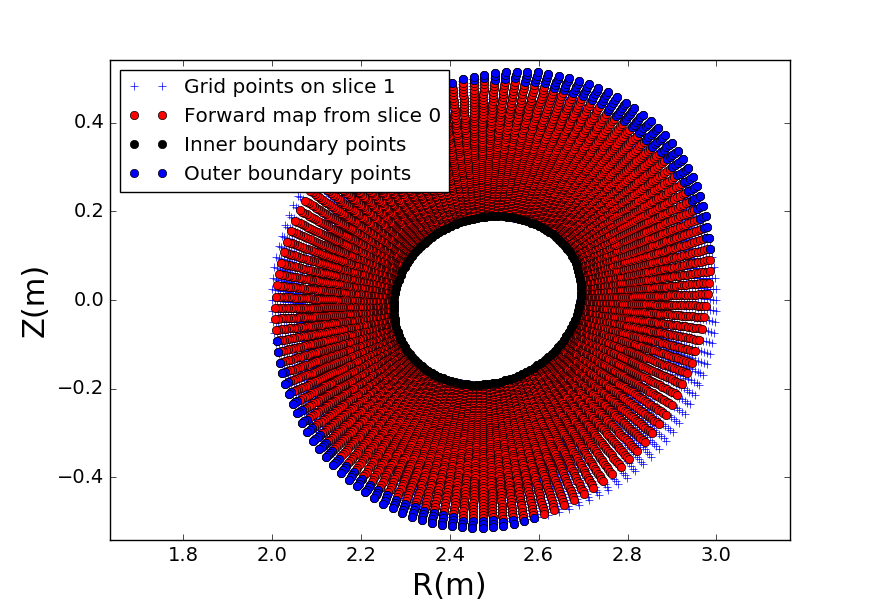}
\caption[]{A curvilinear grid for a rotating ellipse geometry, with an inner surface described by a flux surface, and a circular outer surface providing both open and closed field lines. Blue crosses indicate grid points, whereas circles indicate the locations of field line mapping from the previous plane for the FCI scheme -- red circles indicate field lines which remain in the computational domain, black circles are field lines leaving the inner boundary, and blue circles leave the outer boundary.}
\label{fig:rot_ell_curv_maps}
\end{figure}

Figure~\ref{fig:rot_ell_curv_maps} describes the curvilinear grid used in the following section for simulations of plasma filaments in a rotating ellipse geometry.


This two-field period, rotating ellipse geometry has a major radius of 2.5m, The inner surface is described by a flux surface, but the rest of the grid is not aligned to flux surfaces; the outer surface is a circle centered around the magnetic axis with a radius of 50cm.  Therefore, this geometry incorporates both open and closed field lines.  Figure~\ref{fig:rot_ell_curv_maps} indicates grid points as blue crosses.  The intersection of field lines from the previous plane are indicated by circles: red circles indicate field lines which land within the computational domain, and the remaining circles indicate where the field lines intersect the boundary -- either through the outer surface (blue) or inner (black).  The grid has a resolution of 68x128x16 (radial, poloidal, toroidal), which gives an average poloidal resolution of 0.5cm (radial) by 1.5cm (poloidal).

\section{Nonlinear filament simulations}
\label{sec:filaments}

\subsection{Isothermal Reduced MHD Model}
\label{sec:model}
The following section utilizes a finite-$\mathrm{\beta}$ electromagnetic isothermal reduced magnetohydrodynamic model similar to that used in the isothermal version of TOKAM3X~\cite{Tamain2016} which evolves vorticity $\omega$, electromagnetic potential $A_\parallel$, electron density $n$, and parallel momentum $\Gamma = m_i n v_\parallel$.  Electron and ion temperatures $T_e$ and $T_i$ are assumed constant, though independently specified.  The magnetic field is described by a constant equilibrium field {\bf{$B_0$}} and a time-evolving poloidal field such that:

\begin{align}
  B &= B_0 + \nabla \times \left(A_\parallel e_\phi\right)\\
  &= B_0 + \nabla \psi \times \nabla \phi
\end{align}
where $A_\parallel$ is the parallel component of the vector potential and a large-aspect ratio approximation has been utilized such that $\psi = RA_\parallel$.

The equations are described as follows in SI units:
\begin{align}
  \frac{\partial \omega}{\partial t} + \left({\bf{v_E + v_{\parallel i}}}\right) \cdot \nabla \omega &= \nabla_\parallel{J_\parallel} + \nabla \cdot \left(p \nabla \times \frac{{\bf{b}}}{B}\right) + \nu \nabla_\perp^2 \omega \\
  \frac{\partial }{\partial t} \left[A_\parallel - \frac{m_e}{e}v_{\parallel e} \right] &= -\partial_\parallel \phi + \frac{1}{n} \partial_\parallel p_e - \frac{1}{en}\eta J_\parallel \\
  \frac{\partial n}{\partial t} + {\bf{v_E}}\cdot\nabla n &= -\nabla_\parallel\left(v_{\parallel e} n \right) + \nabla \cdot \left(p_e \nabla \times \frac{{\bf{b}}}{B} \right) \\
  \frac{\partial \Gamma}{\partial t} + {\bf{v_E}}\cdot\nabla \Gamma &=  -\nabla_\parallel\left(v_{\parallel e} \Gamma \right) + \nabla \cdot \left(\Gamma e T_i \nabla \times \frac{{\bf{b}}}{B} \right) - \partial_\parallel p \\
  \omega &= \nabla \cdot \left[\frac{m_i n}{B_0^2} \left(\nabla_\perp \phi + \frac{\nabla_\perp p_i}{en}\right)\right] \\
  J_\parallel &= -\frac{1}{\mu_0} \nabla_\perp^2A_\parallel
\end{align}
Here $\partial_\parallel \equiv {\bf{b}} \cdot \nabla$ and $\nabla_\parallel f \equiv \nabla \cdot \left({\bf{b}}f\right) = B \partial_\parallel \left( \frac{f}{B}\right)$.  The pressure is $p = p_e + p_i = n(T_e + T_i)$.  The vector ${\bf{b_0}} = {\bf{e_\phi}}$ is the ``toroidal'' magnetic field unit vector, and ${\bf{b}} = {\bf{B}}/B_0$ is the unit vector along the total magnetic field, assuming that the poloidal magnetic field is small relative to the toroidal field.  Gradients in the poloidal plane, which is not necessarily perpendicular to the magnetic field (in the case using FCI derivatives, as is used here), are defined by $\nabla_\perp = \nabla - {\bf{b_0 b_0}}\cdot \nabla$.  Dissipation terms are determined by the kinematic viscosity $\nu$ and the resistivity $\eta$, in units of $\mathrm{m^2/s}$ and $\mathrm{\Omega m}$, respectively.

In this model, the magnetic drift term is treated generally (in comparison to, for instance, Equation~\ref{eq:LARcurvature}) and is written as:
\begin{align}
  \nabla \cdot \left[p\nabla \times \frac{{\bf{b}}}{B}\right] &= \nabla \times \frac{{\bf{b}}}{B} \cdot \nabla p \\
  &= \left(\nabla\frac{1}{B^2}\times{\bf{B}} + \frac{1}{B^2}\nabla\times{\bf{B}}\right) \cdot \nabla p \\
  &= -\frac{2}{B^3} \nabla B \times {\bf{B}} \cdot \nabla p \\
  &= \frac{2}{B}{\bf{b}}\times \nabla \log B \cdot \nabla p \label{eq:curvature}
\end{align}
which uses $\nabla \times {\bf{B}} \cdot \nabla p = 0 $ which is valid in equilibrium since ${\bf{J}} \cdot \nabla p = 0$.  The curvature operator is then defined as:

\begin{equation}
  C(f) = \frac{2}{B}{\bf{b}}\times \nabla \log B \cdot \nabla f
\end{equation}
which has a similar form as that derived in the appendix (Equation~\ref{eq:bracket}), meaning that we can use the bracket coefficient to calculate the curvature effects in curvilinear grids.  This is especially convenient as the magnetic field does not, in general, vary solely with the major radius in stellarators -- an approximation which is often used in fluid turbulence simulations~\cite{Shanahan2016, Riva2016, Walkden2016}.  In the simulations presented here, all cross-field drifts are implemented with the 2$\mathrm{^{nd}}$ order Arakawa brackets~\cite{Arakawa1977}.  

\subsection{A weakly-non-axisymmetric, rotating-ellipse geometry}
As an initial investigation of turbulence in non-axisymmetric geometries, a seeded plasma filament in a rotating ellipse geometry was considered.  While there have been experimental investigations of turbulent filaments in stellarators~\cite{Fuchert2016}, this study will serve as the first example of fluid turbulence simulations in non-axisymmetric geometries.  A seeded filament test offers a somewhat straightforward approach to studying important phenomena in plasma transport.  Previous studies in \boutxx~have investigated filaments in slab~\cite{Easy2014}, toroidal pinch~\cite{Riva2016}, and X-point geometries~\cite{Walkden2015, Shanahan2016}.

\begin{figure}[htbp!]
\centering
\includegraphics[width=\textwidth]{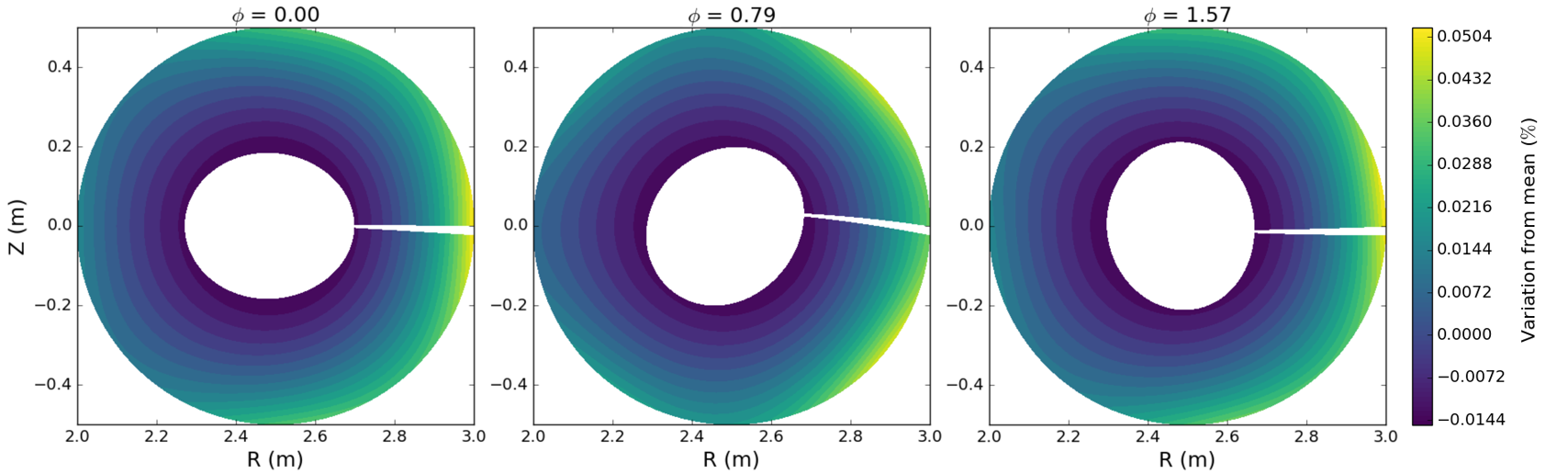}
\caption[]{Variation of the non-toroidal magnetic field at three different toroidal locations -- obtained by multiplying the total field by the major radius R, and calculating the difference with respect to the mean value.}
\label{fig:R*B}
\end{figure}

For the studies presented here, an analytically calculated, low-field-period rotating ellipse geometry was chosen due to the relatively straightforward implementation and analysis.  These analytic equilibria are a necessary step before geometries like W7-X.  Wendelstein 7-X grids for use in BSTING are described in Section~\ref{sec:w7xgrids}, but turbulence studies in these more complex geometries will be a subject of further study.  Furthermore, low-field-period rotating ellipse geometries exhibit a magnetic field which generally varies as $1/R$ (see Figure 1 from~\cite{Loizu2017}), allowing for a more straightforward analysis since this configuration is most similar to axisymmetric configurations. Figure~\ref{fig:R*B} illustrates the degree of non-axisymmetry by plotting the variation of the magnetic field multiplied by the major radius, since a plot of the magnetic field strength would be dominated by the predominantly $1/R$ variation.

From Figure~\ref{fig:R*B} it can be deduced that the magnitude magnetic field which does not vary like $1/R$ only changes toroidally by less than a percent, indicating a small degree of non-axisymmetry in the magnetic field strength, which can affect the drive term for filament propagation (Equation~\ref{eq:curvature}).

\subsection{Filament characterization}
To characterize filament propagation in this non-axisymmetric geometry, a field-aligned plasma filament is first initialized; an approximately circular density perturbation at (R,Z,$\mathrm{\phi}$) = (2.5m,-0.3m, 0.0) is prescribed and a simple parallel diffusion model as in Equation~\ref{eq:diffusion} is first simulated to achieve an initial condition of a field-aligned filament. As this is a low-shear geometry, the filament approximately becomes field aligned once the initial distribution diffuses once toroidally.  The initial field-alignment is determined when the maximum value of the density on a plane varies by less than 5\% in a timestep (100/$\omega_{Ci}$, where $\omega_{Ci}$ is the ion cyclotron frequency).  This condition is satisfied after 100 timesteps, or ten thousand ion cyclotron times.  This field-aligned density distribution, where the peak density perturbation is $n = 1.05\times 10^{19} \mathrm{m^{-3}}$, is then used as an initial condition for the seeded filament simulation using the model described in Section~\ref{sec:model}.  All other plasma fields are not initialized and, once the field-aligned filament is achieved, are allowed to develop independently.  The ion and electron temperature is set to 100eV and the background density is $n_0 = 1\times 10^{19} \mathrm{m^{-3}}$.

Plasma filaments (or blobs) are often characterized by the method by which the charge separation is resolved; if charge is carried via parallel currents through the sheath, filaments are considered ``sheath limited''.  If the connection length to the sheath is large, however, this charge separation can be short-circuited via perpendicular currents and the filaments propagate in a so-called ``inertially-limited'' regime~\cite{Dippolito2011}.  Filament propagation is also characterized by the scaling of the propagation speed as a function of its poloidal cross section, $\delta_\perp$; inertially limited filaments scale proportional to $\delta_\perp^{1/2}$, whereas sheath-limited filaments scale as $\delta_\perp^{-2}$.  For more complete discussions of filaments, see References~\cite{Dippolito2011,Krasheninnikov2008}.  

\begin{figure}[htbp!]
\centering
\includegraphics[scale=.6]{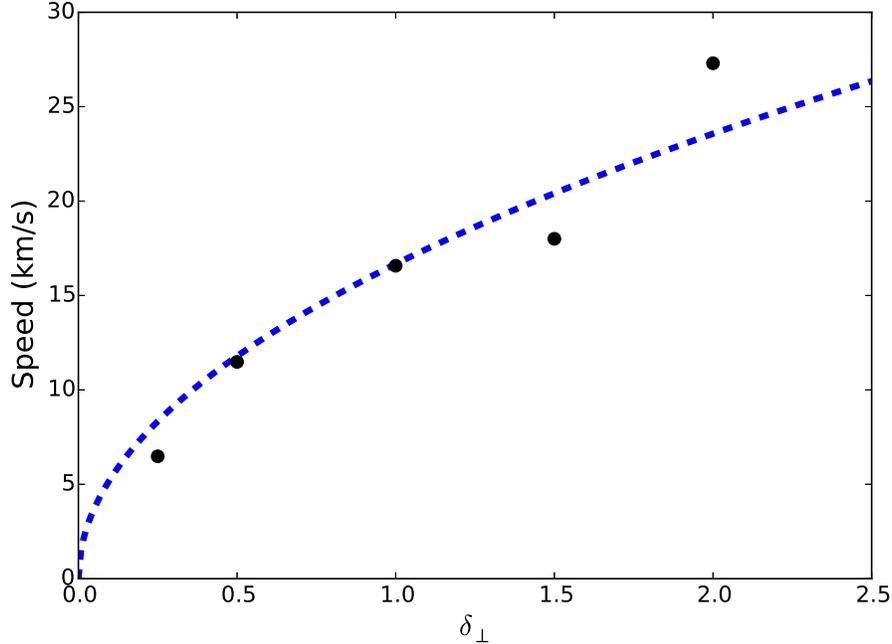}
\caption[]{Inertial filament scaling; filament velocity (circles) and tend to follow a $\delta_\perp^{1/2}$ scaling, indicating propagation in the inertial regime.}
\label{fig:density_scaling}
\end{figure}

Therefore, one can determine the filament propagation regime by plotting the scaling of the maximum speed as a function of filament diameter $\delta_\perp$.  The edge and scrape-off-layer of stellarators such as Wendelstein 7-X can exhibit large connection lengths~\cite{Wurden2017}.  As an initial insight into filament behavior in a non-axisymmetric field with long connection lengths, filaments were seeded in the closed-field-line region in the weakly non-axisymmetric geometry discussed in the previous sections. The scaling of these filaments is shown in Figure~\ref{fig:density_scaling}, where $\delta_\perp = 1$ is normalized to 7cm, the initial filament diameter for the filaments in the following section (\ref{sec:naprop}). 

\begin{figure}[htbp!]
\centering
\includegraphics[scale=.35]{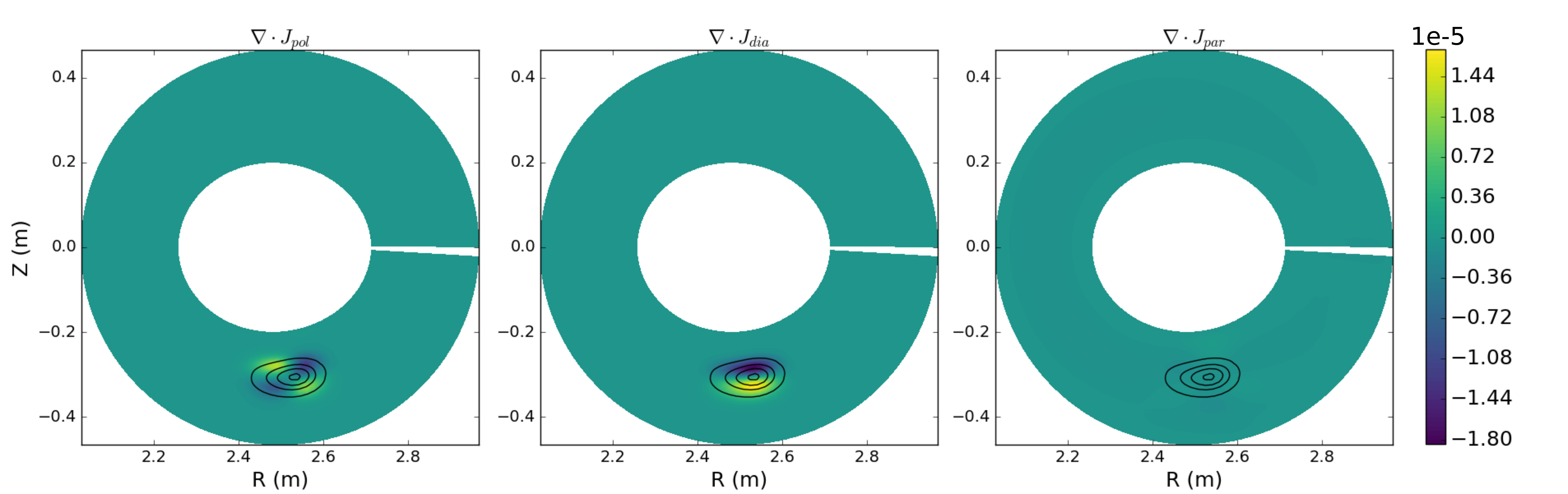}
\caption[]{An illustration of the divergences for parallel and perpendicular currents (color contours) which dictate the propagation of a filament (black contours, overlaid); parallel currents are negligible, indicating inertially-limited propagation.}
\label{fig:Jcomp}
\end{figure}

Similar to the tokamak (axisymmetric) case, the scaling of filaments initialized in the closed-field-line region propagate in an inertially-limited regime, as indicated by the $\delta_\perp^{1/2}$ scaling in Figure~\ref{fig:density_scaling}. As a confirmation of the inertially-limited propagation, Figure~\ref{fig:Jcomp} illustrates the currents which dictate the propagation of the filament at $t \approx 4 \mu s$.

Since the divergence of the parallel current is much smaller than the perpendicular currents, the potential difference is resolved via short-circuiting perpendicular currents, instead of traveling along field lines to the sheath.  This again supports the characterization an inertially-limited regime.  As this is only a weakly-non-axisymmetric field, it is reasonable to find similarities to filaments in an axisymmetric field, for instance in Reference~\cite{Walkden2015}, where inertially-limited filaments were characterized in a MAST (tokamak) geometry.  For a more strongly-non-axisymmetric geometry such as Wendelstein 7-X, the filament propagation may exhibit different behavior, since the filament drive changes directions relative to the major radius within a field period.  While filament simulations in Wendelstein 7-X await a future publication, the following section discusses how even a weakly-non-axisymmetric field can alter the toroidal uniformity of the filament propagation.

\subsection{The effects of nonaxisymmetry}
\label{sec:naprop}
\begin{figure}[htbp!]
\centering
\includegraphics[scale=.5]{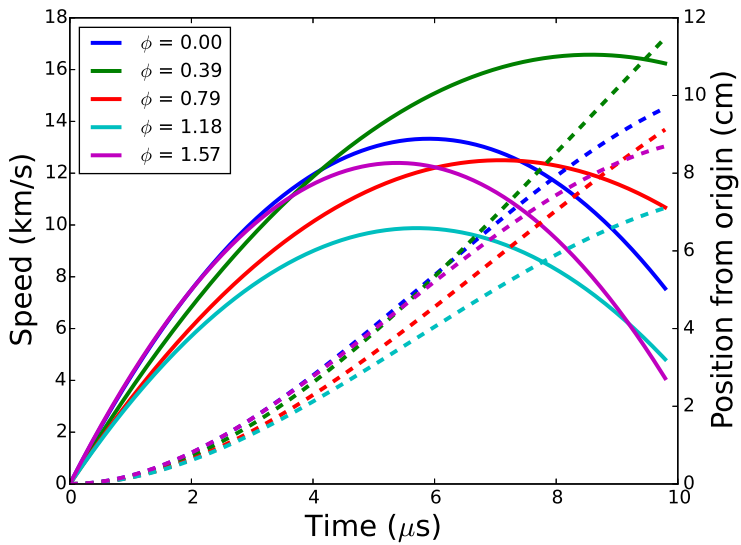}
\caption[]{The effects of nonaxisymmetry; filament velocity (solid) and position (dotted) at various toroidal angles (color) in a rotating ellipse equilibrium}
\label{fig:na_plot}
\end{figure}

If the magnetic geometry is not axisymmetric, the filament drive due to the magnetic field curvature can vary along the length of a filament.  If the drive is toroidally non-uniform, one would expect the propagation to also vary toroidally.  It is often assumed, however, that filaments propagate uniformly along field lines, for instance in~\cite{Fuchert2016}. To test the effects of a non-axisymmetric magnetic field, we can investigate the propagation of a filament at different toroidal locations. Figure~\ref{fig:na_plot} illustrates the filament velocity (solid) and displacement (dotted) of a 100eV plasma filament at various toroidal angles.

\begin{figure}[htbp!]
\centering
\includegraphics[scale=0.5]{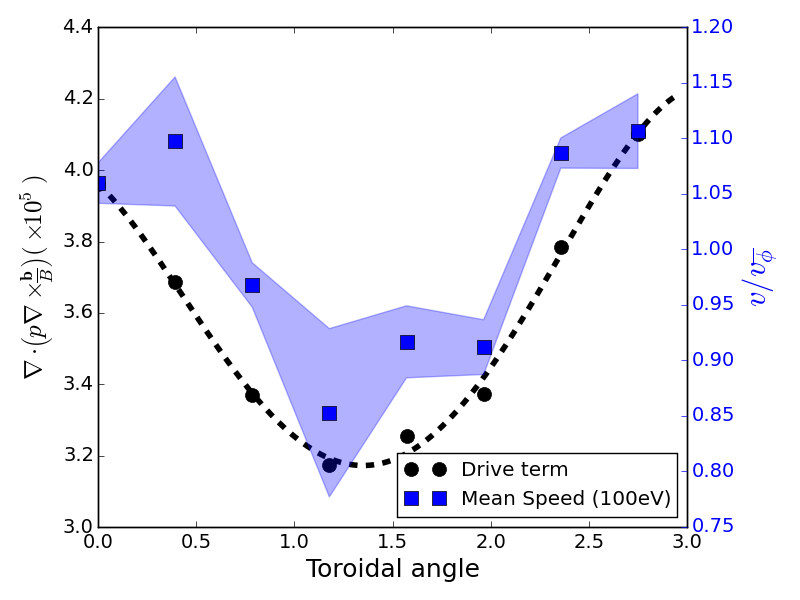}
\caption[]{Weakly non-axisymmetric filament drive, as indicated by the maximum of the magnetic drift term (Equation~\ref{eq:curvature}), representative of the divergence of the diamagnetic current (black).  Also plotted is the time-averaged filament velocity normalized to the toroidally-averaged speed, $v_{\overline{\phi}}$ (blue squares) and the standard deviation (fill) for a 100eV filament.}
\label{fig:NAprop}
\end{figure}

Figure~\ref{fig:na_plot} indicates that even a modestly non-axisymmetric field, as simulated here, can visibly affect the propagation of filaments.  This effect is a direct consequence of the non-axisymmetric filament drive, as shown in Figure~\ref{fig:NAprop} which indicates how the magnetic drive term (black, also fitted), and the resulting filament velocity vary as a function of toroidal angle.  Here, the filament velocity is normalized to the average toroidal field at 100 timesteps.  The blue squares in Figure~\ref{fig:NAprop} indicate the normalized velocity at each toroidal position, averaged over the 100 timesteps.  The fill cloud indicates the standard deviation of the toroidally-normalized velocity for these sample timesteps.  

The non-axisymmetric propagation of filaments can be clarified by considering the timescales associated with filament propagation.  First, we approximate the timescale for parallel propagation along a filament to follow the relation $t_{\parallel} \sim \frac{l}{c_s}$ where $l$ is the length along the filament and $c_s$ is the ion sound speed.  In the simulations presented here, $c_s \approx 6.9 \times 10^{4} \mathrm{m/s}$, which indicates that information takes about 14$\mathrm{\mu s}$ to propagate one meter. Therefore, if the filament is driven non-uniformly, the time which the filament needs to restore the symmetry is longer than the propagation timescale $t_\perp$, which can be approximated by assuming $L_\perp \approx \delta_\perp \approx 7cm$ and $v_\perp \approx 13 \mathrm{km/s}$ -- indicating therefore that $t_\perp \approx 5 \mu s$.

\begin{figure}[htbp!]
\centering
\includegraphics[scale=0.5]{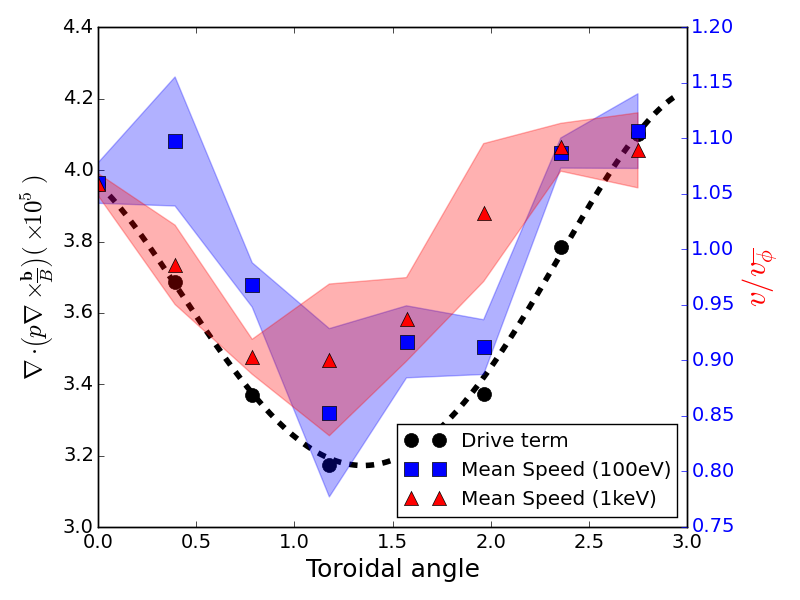}
\caption[]{Filament speed and standard deviation normalized to the average toroidal speed, averaged over 100 timesteps, at each toroidal position for a 100eV filament (blue squares) and a 1keV filament (red triangles).  The more uniform propagation of a hot filament indicates that the sound speed determines the timescale at which non-uniform propagation is mitigated.}
\label{fig:NAprop_hot}
\end{figure}

This assertion can be tested by increasing the speed at which this restoration is performed, for instance by increasing the sound speed.  When simulations were performed with hotter (1keV), smaller filaments -- thus keeping the pressure constant -- the standard deviation of the position of the filaments averaged 79\% of that for the colder simulation, indicating that a hotter filament propagates more uniformly.  This can also be seen in the resulting speed of the hotter filament, shown as red triangles in Figure~\ref{fig:NAprop_hot}, which does not vary as strongly with toroidal location.

\begin{figure}[htbp!]
\centering
\includegraphics[scale=0.5]{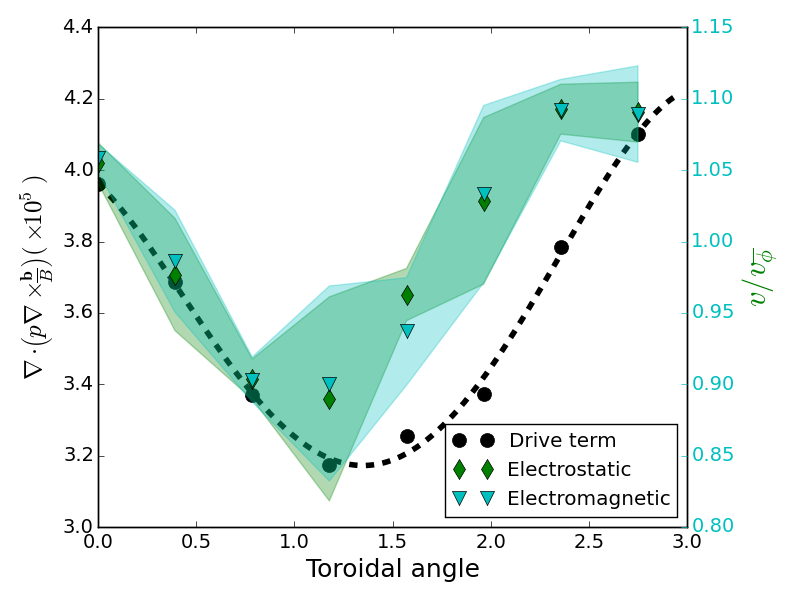}
\caption[]{The non-uniform propagation of an electrostatic (green diamonds) and electromagnetic (cyan triangles) filament as a function of toroidal angle.  Similar propagation indicates that filaments are not restored to uniformity at the Alfv\'en timescale.}
\label{fig:NA_EM_ES}
\end{figure}

It is also possible, however, that the filament is restored to uniform propagation toroidally at the Alfv\'en velocity. This would also explain the more uniform propagation for a hotter filament, since the density perturbation was reduced to provide an equal drive (from pressure), and the Alfv\'en velocity is a function of the plasma $\beta$.  To determine the extent to which this non-axisymmetric nature is affected by the Alfv\'enic effects, one can simulate a filament in an electrostatic case.  In an electrostatic case, all terms in the model described in Section~\ref{sec:model} which are dependent on the plasma $\beta$ are neglected, which in essence provides an infinite Alfv\'en speed.  Figure~\ref{fig:NA_EM_ES} illustrates how the propagation of a filament in an electrostatic and electromagnetic filament compare as a function of toroidal angle.

Figure~\ref{fig:NA_EM_ES} indicates that the non-uniform propagation is not an electromagnetic effect and thus cannot be adequately mitigated by parallel transport at infinite Alfv\'enic speeds, since the electrostatic and electromagnetic case exhibit very similar characteristics.  

\section{Wendelstein 7-X curvilinear grids}
\label{sec:w7xgrids}
BSTING is designed to provide numerical support for experimental measurements.  The curvilinear grid system presented in Section~\ref{sec:poloidalgrids} has therefore been applied to Wendelstein 7-X geometries using various descriptions of the magnetic field.  As this geometry is considerably more complicated than the analytically-prescribed rotating-ellipse equilibria presented earlier, the following sections extend the flux surface mapping tests to the W7-X grids.  

\subsection{Inherent Perpendicular Diffusion in W7-X Curvilinear Grids}
Here we present the development of curvilinear poloidal grids for Wendelstein 7-X geometries using outputs from the VMEC code~\cite{Hirshman1991}.  To test the implementation and limitations of grids in this complicated geometry, the parallel diffusion model in Section~\ref{sec:diffusion}, Equation~\ref{eq:diffusion} was modified to include a perpendicular diffusion, as shown in Equation~\ref{eq:diffusion2}.

\begin{equation}
\label{eq:diffusion2}
\frac{\partial f}{\partial t} = \nabla \cdot \left({\bf{b}}{\bf{b}} \cdot \nabla f \right)  +  D \nabla \cdot \left(\nabla f - {\bf{b_0}}{\bf{b_0}} \cdot \nabla f \right) \equiv \nabla^2_\parallel f   + D \nabla^2_\perp f 
\end{equation}

\begin{figure}[htbp!]
\centering
\includegraphics[scale=.25]{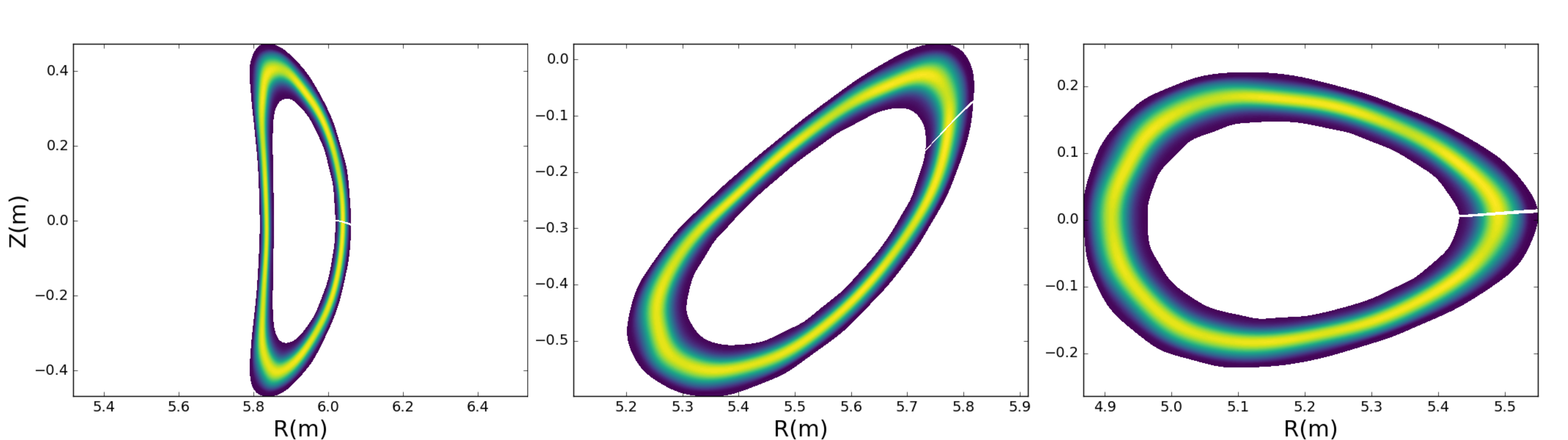}
\caption[]{Three cross sections of the Wendelstein 7-X stellarator indicating flux surfaces as traced by a parallel heat diffusion equation in BSTING}
\label{fig:W7Xfluxes}
\end{figure}

By setting the diffusion coefficient $D$ to zero and simulating Equation~\ref{eq:diffusion2}, we can again recover flux surfaces, similar to the results described in section~\ref{sec:diffusion}. The results of this simulation are shown in Figure~\ref{fig:W7Xfluxes}.

\begin{figure}[htbp!]
\centering
\includegraphics[width=0.6\textwidth]{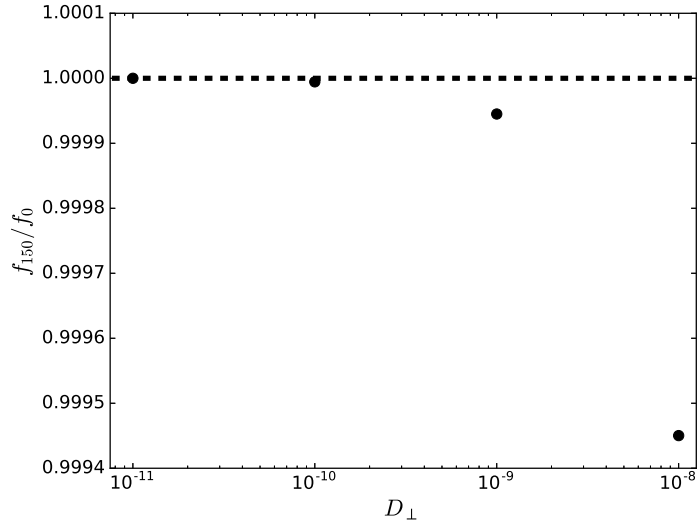}
\caption[]{Proportion of the total test function $f$ at the 150$\mathrm{^{th}}$ timestep normalized to the zero-perpendicular-diffusion case, $f_0$, for several perpendicular diffusion coefficients in a Wendelstein 7-X grid}
\label{fig:W7Xdiffusion}
\end{figure}

Varying the perpendicular diffusion coefficient $D$ allows us to estimate the the inherent perpendicular diffusion in Wendelstein 7-X curvilinear grids. Figure~\ref{fig:W7Xdiffusion} illustrates how the proportion of the test function $f$ at the 150$\mathrm{^{th}}$ timestep compares to the total test function with zero perpendicular diffusion, $f_0$, for various values of $D$ in a Wendelstein 7-X grid with a resolution of 132x16x256 (radial, toroidal, poloidal).  This corresponds to a resolution of approximately 0.3mm -- although this obviously is not uniform -- which is a relatively coarse resolution for a Wendelstein 7-X turbulence study ($\rho_s \approx 0.1mm$).

Figure~\ref{fig:W7Xdiffusion} indicates that the inherent numerical perpendicular diffusion caused by pollution from parallel dynamics is less than a factor of $10^{-9}$ times smaller than the parallel diffusion, as this is where the points begin to diverge significantly from the zero-diffusion case (as indicated by the dashed line at $f_{150}/f_0 = 1.0$).  This inherent perpendicular diffusion is sufficiently less than transport due to plasma drifts and turbulence~\cite{Gunter2005}. This is encouraging as this result is for a moderate-resolution grid, and higher-resolution grids will most likely be necessary for future turbulence simulations in Wendelstein 7-X.  

\subsection{W7-X curvilinear poloidal grid for the edge and scrape-off-layer}
\begin{figure*}[htbp!]
  \centering
  \begin{subfigure}[b]{0.475\textwidth}
    \centering
    \includegraphics[width=0.75\textwidth]{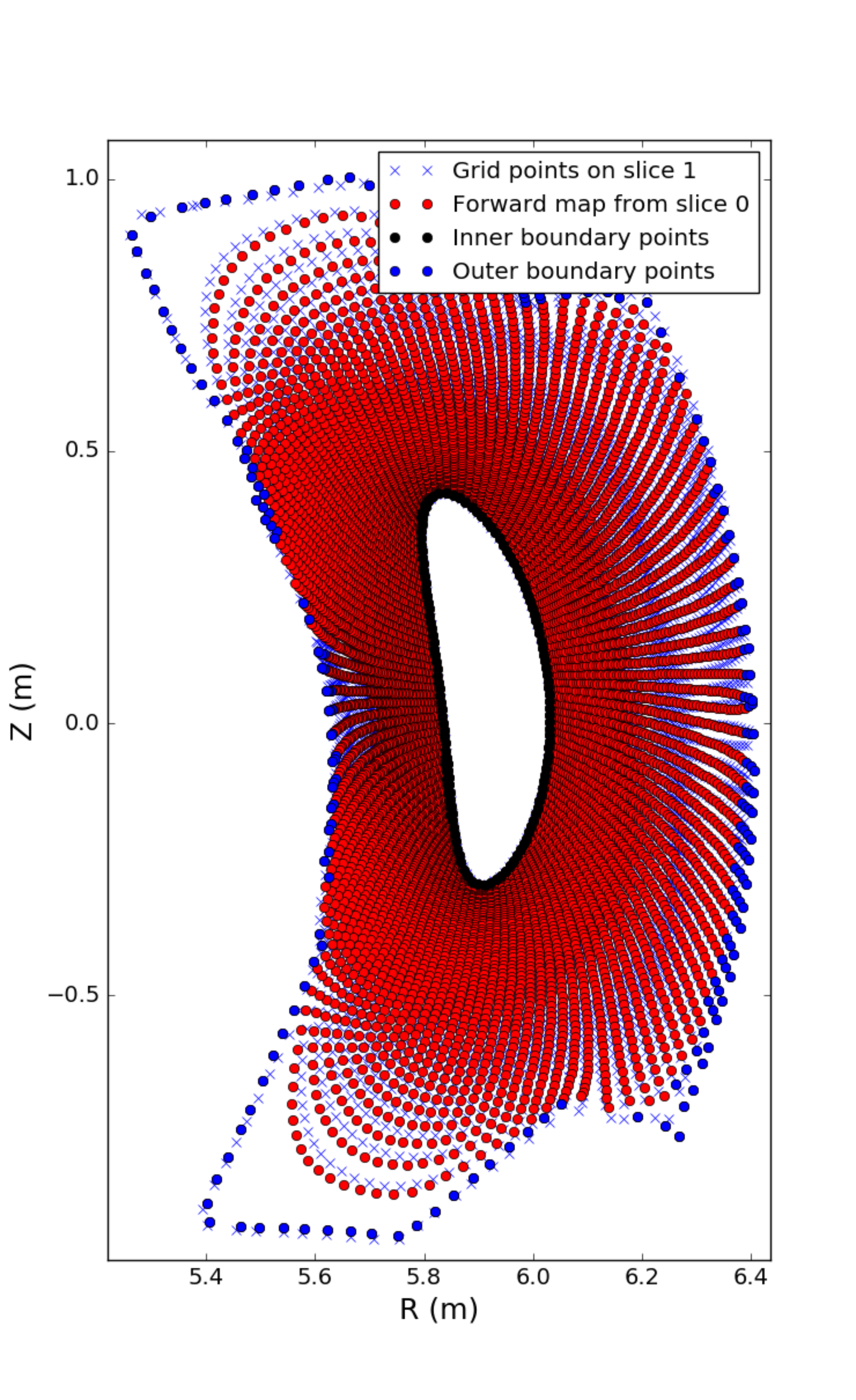}
    \caption[Network2]%
            {{\small Curvilinear grid for the bean-shaped-cross section of W7-X}}    
            \label{fig:w7x_vac_bean_grid}
  \end{subfigure}
  \hfill
  \begin{subfigure}[b]{0.475\textwidth}  
    \centering 
    \includegraphics[width=0.75\textwidth]{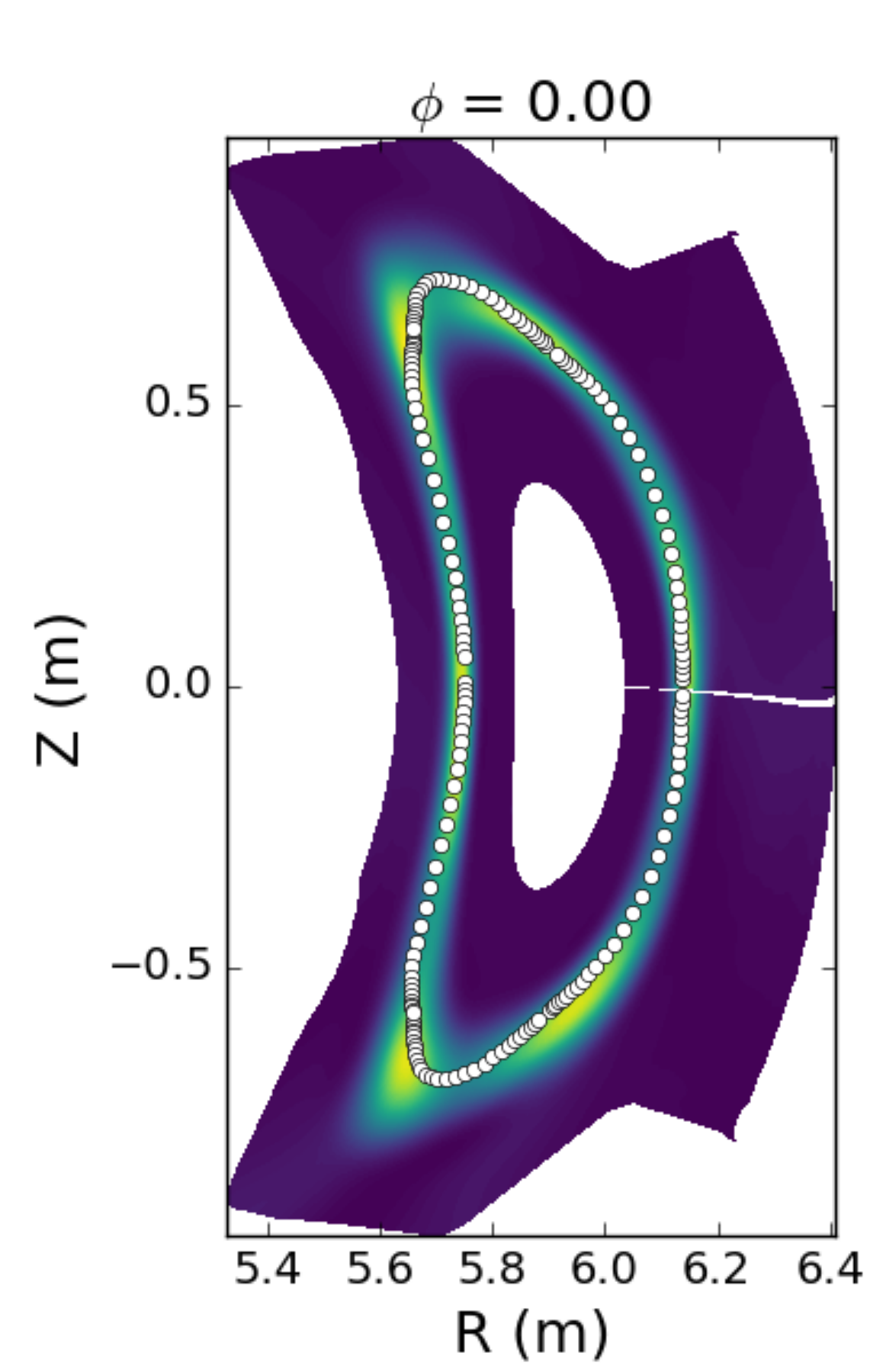}
    \caption[]%
            {{\small Simulated surfaces for the bean-shaped-cross section}}    
            \label{fig:w7x_vac_bean_flux}
  \end{subfigure}
  \vskip\baselineskip
  \begin{subfigure}[b]{0.475\textwidth}   
    \centering 
    \includegraphics[width=0.75\textwidth]{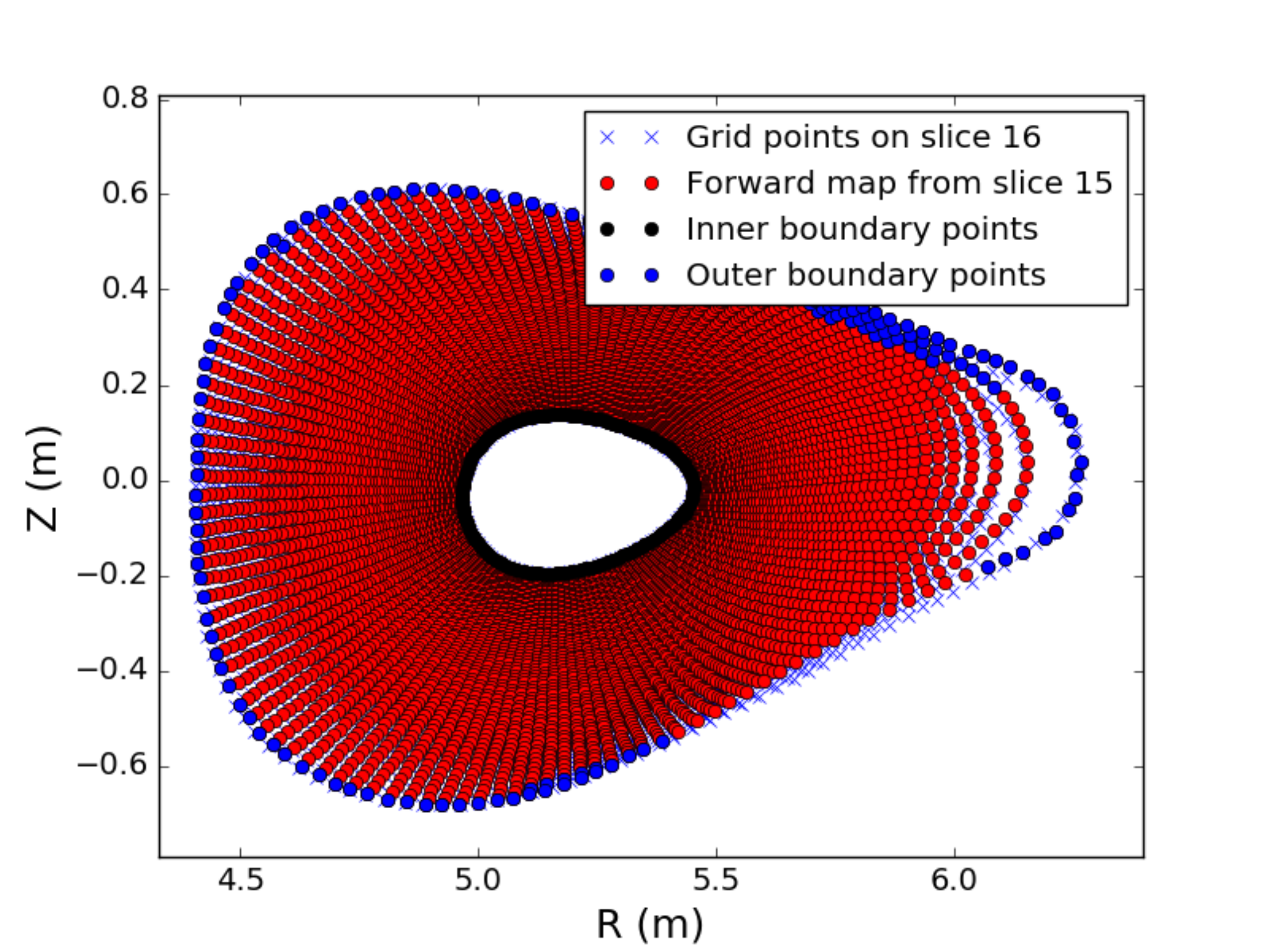}
    \caption[]%
            {{\small  Curvilinear grid for the triangular-cross section of W7-X}}    
            \label{fig:w7x_vac_tri_grid}
  \end{subfigure}
  \quad
  \begin{subfigure}[b]{0.475\textwidth}   
    \centering 
    \includegraphics[width=0.75\textwidth]{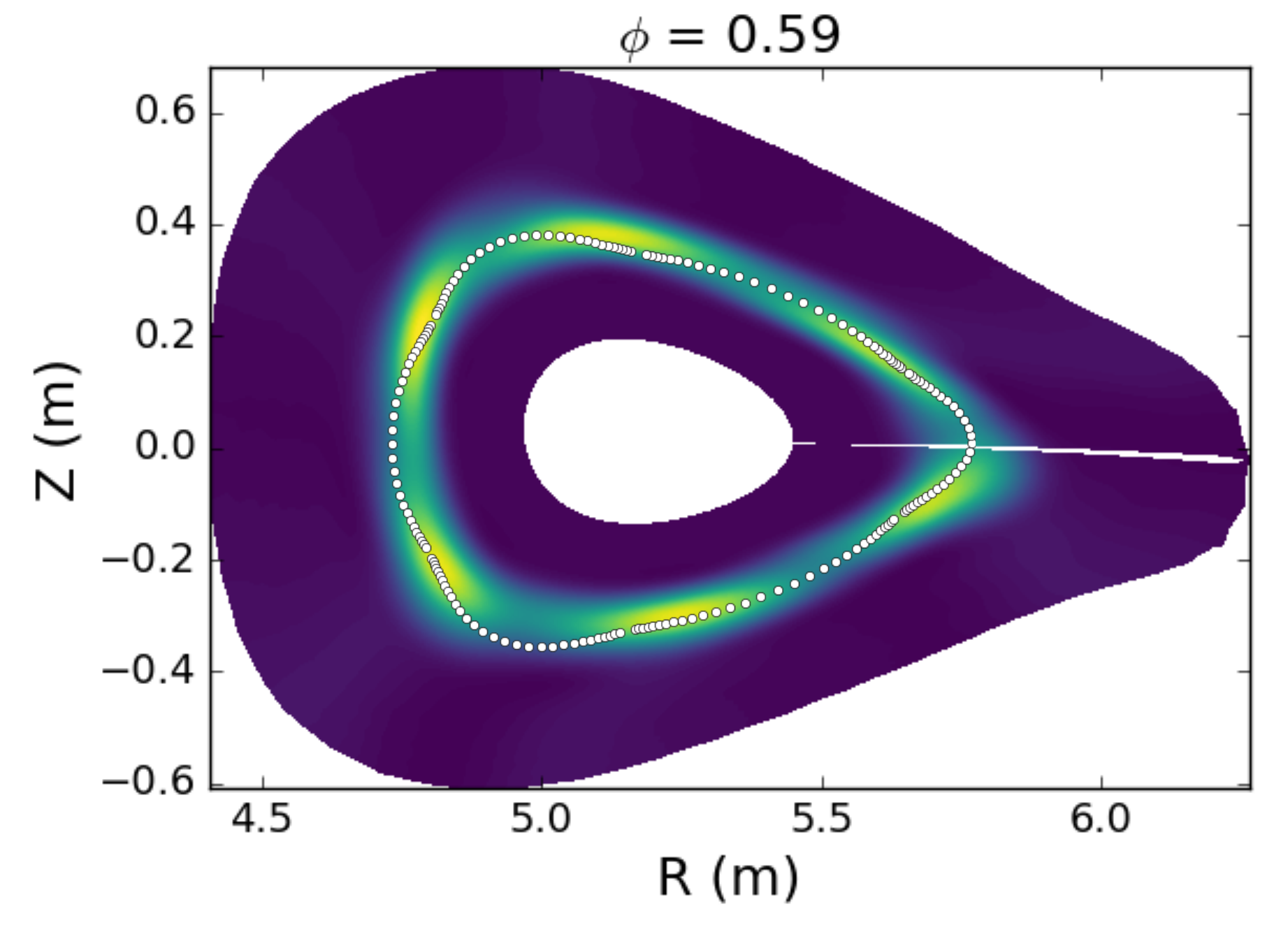}
    \caption[]%
            {{\small  Simulated flux surfaces for the triangular-cross section}}    
            \label{fig:w7x_vac_tri_flux}
  \end{subfigure}
  \caption[ The average and standard deviation of critical parameters ]
          {\small (a,c): Curvilinear grid as generated by the Zoidberg grid generator indicating grid points (blue crosses), and field line maps (circles) for the FCI operators which land inside the domain (red), or leave through the inner (black) or outer (blue) surface.  (b,d): The calculated flux surfaces by parallel heat diffusion simulations for two toroidal locations in W7-X.} 
          \label{fig:w7x_vacuum}
\end{figure*}

The grids described in the previous section are generated from VMEC~\cite{Hirshman1983, Hirshman1991} equilibria, which assume closed flux surfaces.  The edge of Wendelstein 7-X is much more complex as it includes magnetic islands and stochastic magnetic field lines.  As such, another tool must be developed to trace field lines for grids which can accurately describe this region.  To this end, development is ongoing to generate grids based on vacuum field solvers.  Figures~\ref{fig:w7x_vac_bean_grid} and~\ref{fig:w7x_vac_tri_grid} illustrate one such grid, which uses the Wendelstein 7-X web services vacuum field solver and components database~\cite{Bozhenkov2013}.  The inner surface is generated by tracing flux surfaces using a vacuum field solver, which simplifies core boundary conditions and potential coupling to core profiles and sources, and the outer surface is generated based on a description of the Wendelstein 7-X divertor and first wall developed by Michael Drevlak for fast particle calculations, and is also available on the Wendelstein 7-X webservices.


Figures~\ref{fig:w7x_vac_bean_flux} and~\ref{fig:w7x_vac_tri_flux} display the resulting flux surfaces calculated by simulating a parallel diffusion equation on the vacuum curvilinear grid, and overplot an example of a Poincar\'e plot for a nearby flux surface.  While a few challenges remain before full edge simulations of W7-X, this grid serves as a promising first step.  In addition to vacuum field solvers, Zoidberg has also been modified to use EXTENDER~\cite{Drevlak2005}, allowing both plasma-generated magnetic fields and a smooth vacuum solution outside of the last closed flux surface.  

\section{Conclusions}
The \boutxx~framework has been extended to allow the metric tensors to vary in three dimensions.  This provides greater flexibility to the framework.  One major advancement is the implementation of curvilinear grids for use with the Flux Coordinate Independent (FCI) method.  Initial simulations of filament propagation in non-axisymmetric geometry have been performed, and the filaments have been characterized to propagate in the inertially-limited regime. Furthermore, simulations indicate that even a weakly-non-axisymmetric field can significantly alter the propagation of filaments. The long connection lengths of the scrape-off-layer in non-axisymmetric geometries facilitates the establishment of parallel nonuniformity, an effect which must be considered when interpreting experimental data. Since three dimensional effects are becoming increasingly important -- for instance the application of edge magnetic perturbations -- the results presented here are applicable to both tokamak and stellarator configurations. 

Future work will include simulations of filaments in the Wendelstein 7-X stellarator, where the non-uniform drive of a filament can be more pronounced.  The curvature drive in Wendelstein 7-X reverses direction relative to the major radius within a single field period, which could lead to highly non-uniform propagation of filaments, or perhaps even prohibit the radial propagation of coherent filament structures. 


\section{Acknowledgments}
The authors would like to acknowledge the work of the \boutxx~development team. The primary author (BS) would also like to thank Nick Walkden, Joaquim Loizu, and Sophia Henneberg for many useful discussions.

This work has been carried out within the framework of the EUROfusion Consortium and has received funding from the Euratom research and training programme 2014-2018 under grant agreement No 633053. The views and opinions expressed herein do not necessarily reflect those of the European Commission.

\appendix*
\section{Modifications to numerical operators}
 The poloidally curvilinear coordinate system used in this work dictates that numerical operators in the perpendicular (x-z) plane must carefully incorporate the geometry into the calculation. Here we will concentrate on two operators in particular -- Poisson brackets and an example of a curvature operator.
 
 The operator $\frac{1}{B}{\bf{b}}\times\nabla g\cdot\nabla f$ appears often in plasma models and represents phenomena such as \textbf{E}$\times$\textbf{B} advection.  It often appears in equations in the form of Poisson brackets, and is what is referred to here as the bracket operator.  To determine the modifications for the bracket operator in \boutxx, we start by defining real space coordinates $R(x,z)$ and $Z(x,z)$ which depend on the radial coordinate $x$ and the poloidal coordinate $z$.  In the current formulation, $x$ ranges from 0 to 1, and $z$ from 0 to $\mathrm{2\pi}$. From here, we determine the coordinate vectors by taking derivatives along the real-space coordinates:
 \begin{equation}
   {\bf{e_i}} = \frac{\partial}{\partial {\bf{x_i}}} \binom{R}{Z}
 \end{equation}
 where ${\bf{x_i}}$ is either the $x$ or $z$ coordinate. We can now define the metric components as:
 \begin{equation}
   g_{xx} = {\bf{e_x}}\cdot {\bf{e_x}} \quad  g_{xz} = {\bf{e_x}}\cdot {\bf{e_z}} \quad  g_{zz} = {\bf{e_z}}\cdot {\bf{e_z}} 
 \end{equation}
 The $y$-direction is considered to be orthogonal to the $x-z$ plane, and is defined as the toroidal angle spanning 0 to $\mathrm{2\pi}$.  The nonzero metric components are therefore simply:
 \begin{equation}
   g_{yy} = R^2 \quad g^{yy} = \frac{1}{R^2} 
 \end{equation}
 where R is the major radius.
 The unit vector \textbf{b} is considered to be perpendicular to the $x-z$ plane and is defined as:
 \begin{equation}
   {\bf{b}} = \frac{{\bf{e_y}}}{\sqrt{{\bf{e_y \cdot e_y}}}} = \frac{{\bf{e_y}}}{\sqrt{g_{yy}}} = \nabla y \sqrt{g_{yy}}
 \end{equation}
 We can then begin to construct the bracket operator by taking:
 \begin{align}
   {\bf{b}} \times \nabla g &= \sqrt{g_{yy}} \left[\frac{\partial g}{\partial x}\nabla y \times \nabla x + \frac{\partial g}{\partial z}\nabla y \times \nabla z \right]\\
   &= \sqrt{g_{yy}} \left[-\frac{\partial g}{\partial x}\frac{1}{J}{\bf{e_z}} + \frac{\partial g}{\partial z}\frac{1}{J}{\bf{e_x}} \right]
 \end{align}
 Finally, by taking the dot product with $\nabla f$, we get:
 \begin{equation}
   {\bf{b}} \times \nabla g \cdot \nabla f = \frac{\sqrt{g_{yy}}}{J}\left[-\frac{\partial g}{\partial x}\frac{\partial f}{\partial z} + \frac{\partial g}{\partial z}\frac{\partial f}{\partial x} \right]
 \end{equation}
 The terms in the square brackets is defined as the Poisson bracket, which is what is conventionally described in \boutxx~by the bracket operator.  Noting this, we arrive finally at:
 \begin{equation}
   \label{eq:bracket}
   \frac{1}{B}{\bf{b}}\times\nabla g\cdot\nabla f = \frac{\sqrt{g_{yy}}}{JB}\left[g,f\right]
 \end{equation}
where we see that a coefficient of $\frac{\sqrt{g_{yy}}}{JB}$ is required for proper calculation of \textbf{E}$\times$\textbf{B} advection in curvilinear grids.  In Clebsch coordinates, however, it is worth noting that $\nabla z \times \nabla x = \frac{1}{J}{\bf{e_y}}={\bf{B}} $ and therefore $\sqrt{g_{yy}}/J = B$ and this coefficient becomes 1.

 Curvature effects are one of the most important aspects of turbulence simulations, as this can drive drifts and ballooning behavior which contributes to radial transport. The introduction of curvilinear poloidal grids has necessitated careful implementation of curvature operators.  To determine the effects of curvature on a quantity $f$, we must determine how to calculate $\left(\textbf{b}\times\kappa\right)\cdot\nabla f$. As a simple example to illustrate this, we begin by assuming that the curvature vector is of the form:
 \begin{align}
  \kappa &= -\frac{1}{R}\nabla R\\
  &= -\frac{1}{R}\left(\nabla x \frac{\partial}{\partial x} R + \nabla y \frac{\partial}{\partial y} R + \nabla z \frac{\partial}{\partial z} R \right) \\
 \end{align}
 we can then determine:
 \begin{align}
   \left(\textbf{b} \times \kappa \right) &= \nabla y \sqrt{g_{yy}} \times \left(\nabla x \frac{\partial}{\partial x} R + \nabla y \frac{\partial}{\partial y} R + \nabla z \frac{\partial}{\partial z} R \right) \\
  &= - \frac{\sqrt{g_{yy}}}{RJ} {\bf{e_x}} \frac{\partial R }{\partial z}  + \frac{\sqrt{g_{yy}}}{RJ} {\bf{ e_z}} \frac{\partial R }{\partial x}\\
  &= - \frac{\bf{e_x}}{J} \frac{\partial R}{\partial z} +  \frac{\bf{e_z}}{J} \frac{\partial R}{\partial x} \label{eq:LARcurvature}
 \end{align}
which, when dotted with $\nabla f$, then allows the inclusion of curvature effects in curvilinear poloidal grids.  This form of the curvature operator can then be used for large-aspect ratio simulations where the magnetic field varies inversely with major radius, an approximation which is often used in plasma fluid turbulence simulations~\cite{Shanahan2016, Walkden2015}.  A more general curvature operator is derived in Section~\ref{sec:model}.

\tiny{
\bibliography{BSTING}
\end{document}